# Photoconductivity of few-layered *p*-WSe₂ phototransistors *via* multi-terminal measurements

Nihar R. Pradhan [a,*], Carlos Garcia [a,b], Joshua Holleman [a,b], Daniel Rhodes [a,b], Chason Parker [a,c], Saikat Talapatra [d], Mauricio Terrones [e,f], Luis Balicas [a,*] and Stephen A. McGill [a,*]

[a]National High Magnetic Field Laboratory, Florida State University, Tallahassee, FL 32310, USA
[b]Department of Physics, Florida State University, Tallahassee, FL 32306, USA
[c]Leon High School, Tallahassee, FL 32308, USA
[d]Department of Physics, Southern Illinois University, Carbondale, IL 62901, USA
[e]Department of Physics, Pennsylvania State University, PA 16802, USA
[f]Center for 2-Dimensional and Layered Materials, Pennsylvania State University, PA 16802, USA

**Abstract:** Recently, two-dimensional materials and in particular transition metal dichalcogenides (TMDs) were extensively studied because of their strong light-matter interaction and the remarkable optoelectronic response of their field-effect transistors (FETs). Here, we report a photoconductivity study from FETs built from few-layers of *p*-WSe₂ measured in a multi-terminal configuration under illumination by a 532 nm laser source. The photogenerated current was measured as a function of the incident optical power, of the drain-to-source bias and of the gate voltage. We observe a considerably larger photoconductivity when the phototransistors were measured *via* a four-terminal configuration when compared to a two-terminal one. For an incident laser power of 248 nW, we extract 18 A/W and ~4000% for the two-terminal responsivity (*R*) and the concomitant external quantum efficiency (*EQE*) respectively, when a bias voltage $V_{ds}$ = 1 V and a gate voltage $V_{bg}$ = 10 V are applied to the sample. *R* and *EQE* are observed to increase by 370% to ~85 A/W and ~20000% respectively, when using a four-terminal configuration. Thus, we conclude that previous reports have severely underestimated the optoelectronic response of transition metal dichalcogenides, which in fact reveals a remarkable potential for photosensing applications.

## Introduction

Two-dimensional van-der Waals coupled crystals, particularly transition metal dichalcogenides (TMDs), display a promising potential for applications in electronics and optoelectronics due to their inherently strong interaction with light and a band gap that lies in the visible region [1, 2]. Among the TMDs, MoS₂ has attracted a lot of attention due to its natural availability and its electrical and optical properties, namely a direct band gap of 1.8 eV when in single atomic layer form [3], its excellent FET behavior with high ON/OFF current ratio of ~10⁸ [4, 5], and a high responsivity when in single atomic layer form.

Previous reports based on monolayer or on few-layered MoS$_2$ show wide ranging responsivities, i.e. from few mA/W to $10^4$ A/W [5-8], depending upon the applied drain to source bias, gate voltage, optical power and the quality of the electrical contacts. Sanchez *et al.*, [5] reported an extremely high responsivity from monolayer MoS$_2$ that approached ~880A/W under an applied gate voltage $V_g$ = 60V and with a drain-source voltage $V_{ds}$ = 8 V. Z. Yin *et al.*, [6] demonstrated an increase in responsivity at $V_g$ > 0 V for monolayer MoS$_2$ when compared to the OFF state responsivity of the transistor at $V_g$ < 0 V. In contrast to these observations indicating higher responsivity in monolayer MoS$_2$ when $V_g$ > 0 V (or in the ON state of the transistor), Pak *et al.* [8] and Lee *et al.* [9] reported high responsivities in few-layered MoS$_2$ but when $V_g$ < 0 V or when the FET is in its OFF state. These studies suggest that the sensitivity of the photoconductivity and related transport mechanism is different between monolayer and few-layered materials. Similar contrast in the photoresponse between the ON and the OFF states of the transistor were also reported for CVD grown monolayer MoSe$_2$ crystals [10].

Similar to MoS$_2$, WSe$_2$ is also a layered semiconductor and has a band gap ranging from 1.3 eV in the bulk to 1.8 eV in a single atomic layer but its electro-optical properties have been less studied. Unlike MoS$_2$ and MoSe$_2$, recent reports indicate that WSe$_2$ exhibits *p*-type conduction [11]. Phototransport studies on monolayer WSe$_2$ show a responsivity of ~8 mA/W when the transistor is in its OFF state of operation. This responsivity increases drastically when the FET is in the ON state under an applied gate voltage $V_g$ = -60 V and a bias voltage $V_{ds}$ = 2 V, albeit showing a very poor switching speed of ~5 s [12]. This study also claims that the contact area between the metallic electrodes and the semiconductor plays an important role for the photoconductivity. When Ti instead of Pd was used for the contacts [12], leading to a higher Schottky barrier, the switching speed decreased to just 23 ms. However, the photogain also decreased by several orders of magnitude. In contrast, our previous results on tri-layered WSe$_2$ crystals [13] revealed a much smaller photoresponse time of just ~5 μsec, which is $10^6$ times faster than those measured from samples based on CVD grown monolayers [12]. This could be attributed to the higher crystallinity of the samples produced by the CVT synthesis technique which leads to a lower density of defects, and also to the multi-layered nature of the WSe$_2$ crystal used.

Recent and past photoconductivity and electrical transport studies on TMDs and other materials performed *via* a two-terminal configuration of contacts reveal a potential for optoelectronic applications. Their phototransport properties can be tuned by the materials used for the electrical contacts [13], doping [14, 15] and also by the environment making them suitable for gas sensing applications [16]. The carrier mobility is dominated by the resistance of the contacts, which limits the performance of the transistors [11, 17-19]. Recently, a number of reports claimed that intrinsically higher carrier mobilities are observed when using a four-terminal contact configuration, which, in contrast to the two-terminal measurements, diminishes the influence of the resistance of the contacts on the electrical transport properties [17, 19]. To date, all reports on the optoelectronic properties of two-dimensional materials used a two-terminal instead of a multi-terminal configuration of electrical contacts which should reduce the influence of the contacts. In this study we demonstrate that by using a multi-terminal configuration for the contacts one extracts significantly higher values for the photoconductivity of few-layered $WSe_2$ phototransistors precisely by minimizing the role of the resistance of contacts. The responsivity and the external quantum efficiency of few-layered $WSe_2$ phototransistors measured *via* a four-terminal configuration is found to be ~370% higher than the respective values measured through a two-terminal one on the same device. Therefore, our results suggest that the intrinsic responsivity of transition metal dichalcogenides are likely to be considerably higher than the values previously reported, thus opening interesting prospects for the development of phototransistors, photodiodes or photosensors based on these compounds.

$WSe_2$ single-crystals were grown *via* a chemical vapour-transport (CVT) technique using iodine as the transport agent. The details of the synthesis procedure can be found in our previous reports and also in experimental methods [6, 8, 20]. These crystals were characterized through photoluminescence (PL), Raman, and electron diffraction (EDX) spectroscopies to confirm the composition and the high quality of the single-crystals [11, 20]. Few-layered $WSe_2$-FETs, were fabricated using standard electron beam lithography techniques, where thin layers of $WSe_2$ were mechanically exfoliated and subsequently transferred onto a 270 nm thick $SiO_2$ dielectric substrate grown on a *p*-doped Si wafer. The electrical contacts, i.e.Ti/Au (5/80 nm), were deposited via e-beam evaporation. Figure 1 (a) presents the schematic of a WSe$_2$ phototransistor on Si/SiO$_2$ substrate along with the measurement scheme. Figure 1 (b) corresponds to the optical micrograph of one of the samples having six contacts which allows both two-terminal (2-terminal) and four-terminal (4-terminal) measurements. We used two voltage leads ($V^+$ and $V^-$) and two current leads ($I^+$ and $I^-$), to measure the intrinsic photoconductivity of the device (Fig. 1 (b)), while only the two of the leads ($I^+$ and $I^-$) were used to evaluate the phototransport properties in a two-terminal configuration [17,19]. We used Keithley sourcemeters, models 2400 and 2612A, to source the voltage, sense current, and apply the gate voltage to our samples. The back gate voltage ($V_{bg}$) was applied to the highly $p$-doped Si substrate to accumulate carriers in the channel. Figure 1 (c) presents an atomic force microscope (AFM) topography profile across the edge of the device shown in Fig. 1 (b) which was collected along the dotted blue line. The thickness indicates that the WSe$_2$ crystal consists of ~10 atomic layers. The channel length and its average width are $l$ = 15.8 μm and $w$ = 7.7 μm, respectively. Results from this device will be discussed in detail throughout this manuscript.

**Results and discussion**

We evaluated the electrical transport properties of few-layered WSe$_2$ field-effect transistors fabricated on a Si/SiO$_2$ substrate using $p$-doped Si as the back-gate electrode. All measurements were performed under ambient conditions in a dark room environment. Figure 2 (a) displays the source-drain current ($I_{ds}$) measured in a 2-terminal configuration as a function of the source-drain voltage ($V_{ds}$) under several values of the applied gate voltage ($V_{bg}$). Figure 2 (b) shows the equivalent transfer characteristics of this device but measured *via* a 4-terminal configuration. This data is plotted on a semilogarithmic scale in supplementary Fig. S2. Two-terminal measurements were performed through the two contacts at the extremities of the sample which were used to both source the voltage and measure the current. The same contacts were used to inject the current for a four-terminal configuration of measurements, with the two inner contacts used to sense the voltage, as depicted in Fig. 1 (b) [20].

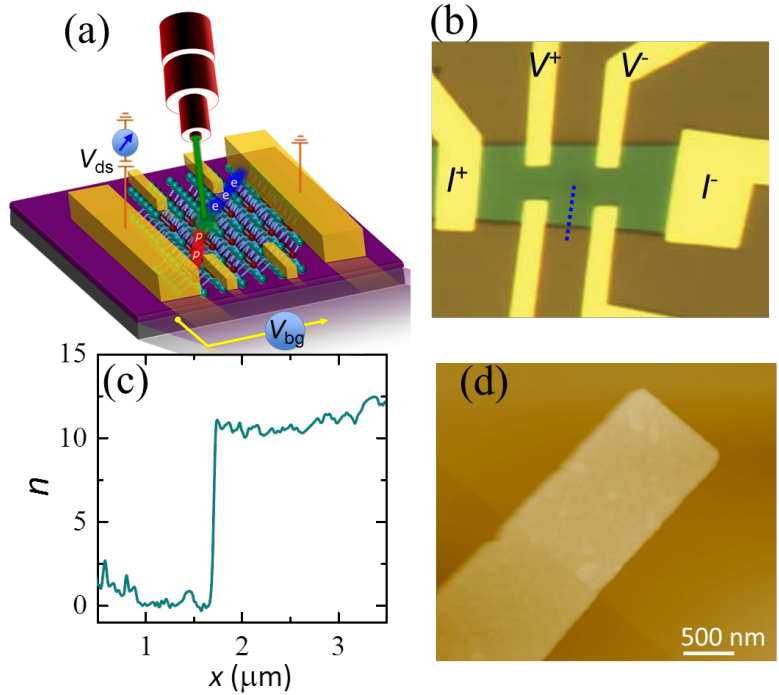

**Fig. 1** (a) Schematic of a WSe$_2$ phototransistor on a Si/SiO$_2$ substrate and scheme for photoconductivity measurements. (b) Optical image of a four-terminal transistor fabricated on a SiO$_2$ substrate. Highly *p*-doped Si was used as the back gate. The channel length and width are $l$ = 15.8 μm and $w$ = 7.7 μm, respectively. (c) AFM height profile of the WSe$_2$ crystal indicates that it is composed of 10 atomic layers (monolayer thickness ~0.7 nm) and (d) AFM image of the flake shown in (b).

It is clear from the transfer characteristics, shown in Figs. 2 (a) and 2 (b), that the 4-terminal measurements show much higher currents when compared to 2-terminal measurements, which is attributable to the elimination of the contact resistance which affects the 2-terminal measurements [19, 20]. The $I_{ds}$ as a function of $V_{ds}$ traces show linear behavior which is due to thermionic emission of carriers above the Schottky barrier between the Ti/Au contacts and the WSe$_2$ crystal. Figures 2 (c) and 2 (d) show the $I_{ds}$ as a function of the applied back-gate voltage $V_{bg}$ for several constant values of $V_{ds}$ measured with both the 2-and the 4-terminal configurations. From Figs. 2 (c) and 2 (d) we can see a modest electron current at positive gate voltages but the drain-source current increases considerably under negative gate voltages, indicating that our CVT grown WSe$_2$ crystals are hole-doped. The maximum current was limited by fixing the maximum value of the applied drain to source voltage in order to prevent damage of our samples. The two-terminal mobility of the charge carriers in our WSe$_2$ FETs can be deduced from the linear slope of $I_{ds}$ as a function of $V_{bg}$, see Fig. 2 (c) and also supplementary Fig. S3 (a), through the following expression:

$$\mu = \left(\frac{dI_{ds}}{dV_{bg}}\right) \times \left(\frac{L}{WV_{ds}C_i}\right) \qquad (1)$$

where $C_i = \varepsilon\varepsilon_o/d = 11.783\times10^{-9}$ F/cm$^2$ is the gate capacitance per unit area, $\varepsilon$ (= 3.9) is the dielectric constant of SiO$_2$, $\varepsilon_o$ is the dielectric constant of vacuum and $d$ is the thickness of the 278 nm thick SiO$_2$ layer, $L$ is the channel length (15.8 µm) or the distance between the two current contacts and $W$ (=7.7 µm) is the width of the channel. The two-terminal field-effect *hole-mobility* of the device calculated from (1) is ~220 cm$^2$/Vs. The four-terminal field-effect *hole-mobility* of the device can be calculated using the following expression [17, 19]:

$$\mu = \left[\frac{l_v}{wC_i}\right] \times \left[\left(d\left(\frac{I_{ds}-I_0}{V_{12}}\right)/dV_{bg}\right)\right] \qquad (2)$$

where $l_v$ is the length between the two voltage leads, $I_0$ is the off-current, and $V_{12}$ is the voltage measured between the voltage leads. The four-terminal field-effect hole-mobility of the device calculated from (2) is 145 cm$^2$/Vs. Our four-terminal mobility is very close to the value previously obtained via our Hall-effect measurement [11]. The measured Hall capacitance in our device is about 1.5 times larger than the estimated capacitance using a 270 nm thick SiO$_2$ layer. This is likely due to impurities or defects within the SiO$_2$ layer which cause the capacitance to be higher than the estimate. The difference in field-effect mobilities we report for the two- and four-terminal data may be due to an inhomogenous accumulation of charge within the channel. The channel lengths for the two- and four-terminal measurements were 15.7 µm and 6.1 µm, respectively. It is possible therefore that our two-terminal mobility represents more of a "global" measure of the channel field-effect mobility. A more systematic approach for future work might be to sample the four-terminal transconductance at various points within the channel. This could lead to an average four-terminal mobility more consistent with the two-terminal measurement. Others have also reported similar differences in four- and two-terminal mobilities for MoS$_2$ and InSe FETs [21, 22].

In addition, our few-layered WSe$_2$ FET exhibits a high ON to OFF current ratio of ~$10^6$ (see, Supplementary Fig. S3 (b)). The presence of defects or trap states between the SiO$_2$ layer and the semiconductor plays a detrimental role in the mobility of the device; therefore the mobility could be improved by using a suitable substrate such as clean hexagonal boron nitride (*h*-BN), or another substrate characterized by a low density of charge traps. By comparing Figs. 2 (c) and 2 (d) it becomes clear that the current measured when using a 4-terminal geometry is considerably higher than the 2-terminal one due to the minimization of the role of the resistance of the contacts. Nonetheless, the extracted current is still affected by the size of the Schottky barrier at the contacts. The calculated contact resistance is given in the supplementary Fig. S3 (b). Our aim here is to evaluate the photoconductivity using a multi-terminal geometry for the contacts and to compare it with the values extracted when using a conventional two-terminal configuration, and in this way evaluate the near intrinsic photoresponse of these compounds.

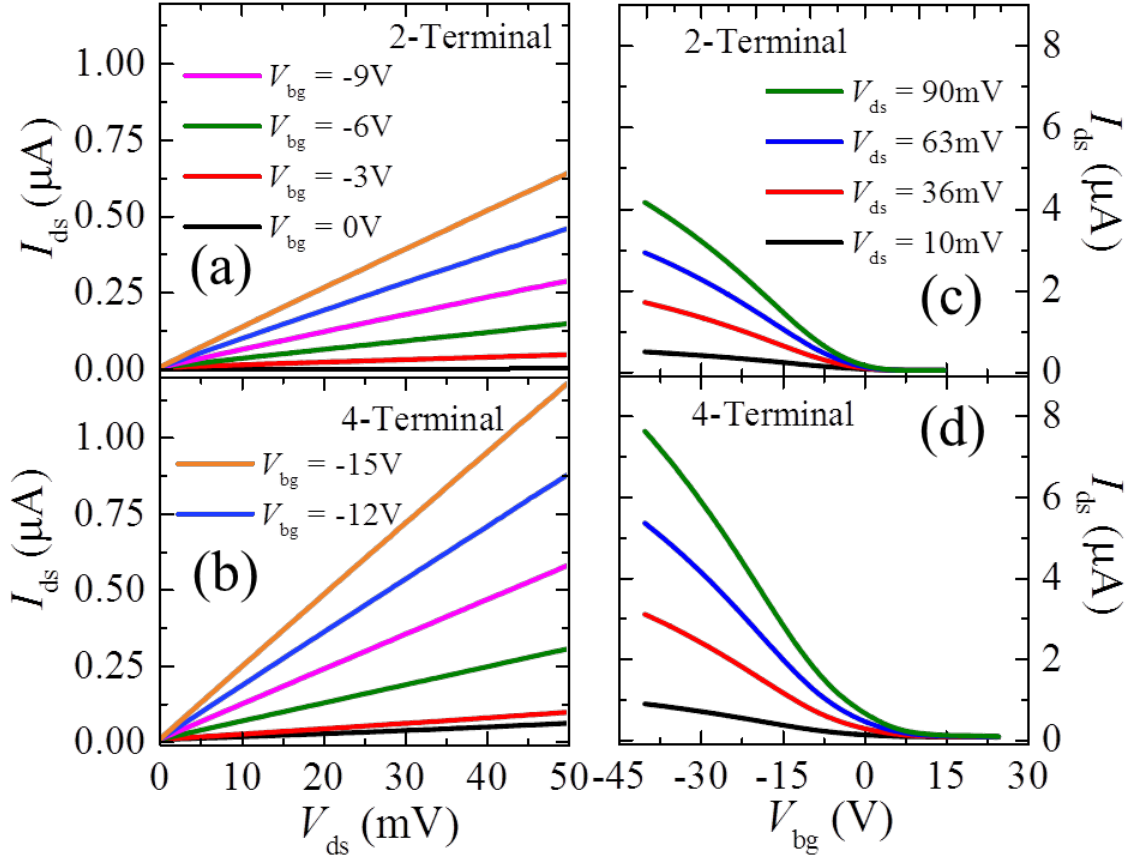

**Fig. 2** Comparison between the transport properties of few-layered WSe$_2$ FETs when measured in a 2- and in a 4-terminal configuration. (a) and (b) Drain to source current ($I_{ds}$) as a function of the drain-source voltage ($V_{ds}$) in two- and in four-terminal configuration, respectively under several back gate voltages ranging from 0 to -15V. (c) and (d) $I_{ds}$ as a function of the applied gate voltage $V_{bg}$ for 2- and 4-terminal configurations for several drain-source voltages.

The transport data presented in Fig. 2, shows that the OFF current ($10^{-11}$ A to $10^{-10}$ A) varies little for either configuration of measurements, i.e. for either the 2-terminal or 4-terminal configurations. This is probably due to the depletion of the charge carrier in the conducting channel. Hence, the measured dark current (without illumination), in the OFF state of the transistor does not significantly change in 4-terminal when compared to the 2-terminal configuration of measurements.

Photoconductivity measurements were performed under laser illumination with a wavelength of 532 nm coupled to a home-built microscope and data acquisition system. A sketch of the experimental setup is shown in Fig. 3 below. The output power of the laser was controlled through a variable attenuator, focused with a lens and subsequently transmitted through an optical fiber. The signal was split with 50/50 beamsplitter to monitor half of the incident power while the other half was sent to the sample.

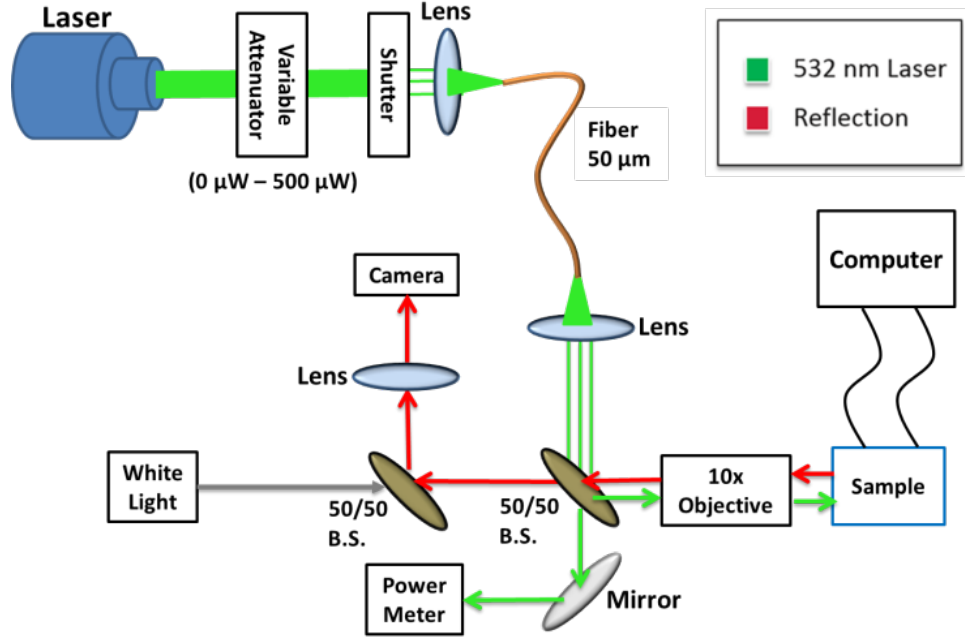

**Fig. 3** Optical setup for our phototransport characterization of $WSe_2$ phototransistors. A variable attenuator was used to continuously vary the output laser power incident on the sample. White light is used for aligning the laser spot.

Since there is a limited amount of charge carriers in the OFF state of the transistor, we focused on the photoconductivity in this region. After aligning the laser spot onto the sample, we calibrated the illumination power $P_{opt}$ irradiated onto it which is controlled by the attenuator. Here, $P_{opt} = \frac{P}{\pi r^2} \times A$, where $P_{opt}$ is the optical power illuminating the sample area, $P$ is the total laser power, $r$ is the radius of the spot size, and $A$ is the area of the device). Initially, we recorded the dark current $I_{dark}$, in absence of illumination, as a function of $V_{ds}$ ($I_{dark} = I_{ds}$ in absence of illumination) ranging from 0 to 1 V under $V_{bg}$ =10 V, or when the transistor is completely in its OFF state. As a subsequent step, we measured the drain-source current, $I_{ds}$, under illumination of the entire channel area using the 532 nm laser. We collected $I_{ds}$ by continuously sweeping the laser power by using a variable attenuator in front of the laser as shown in Fig. 3. The diameter of the laser spot is larger than our sample size and was measured with lithographically patterned grids on a Si wafer to calculate the power density/power illuminated onto the sample area. With our optical setup we swept the optical power from 248 nW to 100 µW and measured the drain-source current at fixed values of both $V_{ds}$ and $V_{bg}$. The photogenerated current ($I_{ph}$) was extracted by subtracting the dark current ($I_{ph} = I_{ds} - I_{dark}$) measured under the exact same value of $V_{ds}$.

Figure 4 displays the measured photocurrent measured via both 2- and 4-terminal configurations under a fixed gate voltage $V_{bg}$ = 10 V. Figure 4 (a) displays the 2-terminal responsivities ($R = I_{ph}/P_{opt}$) in a logarithmic scale as a function of the laser power illuminating the sample area and for several values of $V_{ds}$ ranging from 10 mV to 1 V. All measurements were performed under an applied gate voltage $V_{bg}$ = 10 V. The photocurrent gradually increases as a function of the laser power for both 2- and 4-terminal measurements. The photocurrent as a function of the laser power when measured in a 4-terminal configuration is shown in the electronic supplementary information (Fig. S1). In a logarithmic scale the responsivity is seen to decrease linearly as a function of the illumination power, which can be explained in terms of trap states present either in the $WSe_2$ or at the interface between $WSe_2$ and the $SiO_2$ substrate. Trap states present in $WSe_2$ and/or at the interface can have a significant impact on the dynamics of photodetectors based on $WSe_2$ [5, 23]. We have not studied the photoresponse time in this report as we focused only on photoconductivity of $WSe_2$ in multi-terminal configurations. Recently we reported the time resolved measurements on tri-layered $WSe_2$ indicate that the characteristic decay/raise time after turning the laser either ON or OFF can be as low as $\tau$ ~ 5 - 10 µsec [13]. A similar low photoconductivity response time ~5 - 12 µsec was also observed on few-layered $SnS_2$ [24] using acoustic-optic modulator as the light chopping method. The response time was observed to be slower in the study using the mechanical chopping method [24]. This value of response time measured in acoustic modulation is six orders ($10^6$) of magnitude faster than the reported characteristic photoconductivity decay time of photodetectors based on monolayer $MoS_2$ [5] and three orders of magnitude ($10^3$) faster than the characteristic response time of a $WSe_2$ monolayer photodetector [12]. The highest responsivity obtained from our few-layered $WSe_2$ transistors measured *via* a two-terminal configuration is 18 A/W for $V_{ds}$ = 1 V and $V_{bg}$ = 10 V when the transistor is in its OFF state and under an illumination power $P_{opt}$ = 248 nW

(Fig. 4 (a)). When we extract the responsivity from a 4-terminal configuration, the corresponding value of responsivity increases up to ~ 85 A/W (Fig. 4 (c)), which is nearly 370% higher than the value obtained from the 2-terminal configuration. We fitted the responsivity as function of $P_{opt}$ to a power law. From this fitting ($R \propto P^{\gamma}$), we obtained the exponent $\gamma \sim 0.98$. From the above responsivity, we determined the external quantum efficiency (*EQE*), which is defined as the number of electron-hole pairs generated by the number of incident photons onto phototransistor. *EQE* was estimated from the following equation:

$$EQE = \frac{I_{ph}}{P_{opt}} \times \frac{hc}{\lambda q} = R \times \frac{hc}{\lambda q} \qquad (3)$$

Where $h$ is the plank constant, $c$ is speed of light, $\lambda$ is the wavelength of the laser source used, $P_{opt}$ is the incident optical power and $q$ is the electron charge. The values of *EQE* obtained from our few-layered WSe$_2$ phototransistor as a function of $P_{opt}$ measured respectively in 2- and 4-terminal configurations are displayed in Fig. 4 (b) and (d), respectively. *EQE* decreases as a function of the excitation power but increases as a function of the applied $V_{ds}$. When using a two-terminal configuration, the lowest observed *EQE* value under $P_{opt}$ = 248 nW and $V_{ds}$ = 10 mV is 49% which increases to 152% when a 4-terminal one is used (see Figs. 4 (b) and (d)). For $V_{ds}$ = 1 V the highest *EQE* value measured in a 2-terminal configuration is ~4500% which increases up to ~20000% when measured with a 4-terminal one (see, Figs. 4 (b) and (d)). Thus, in our few-layered WSe$_2$ phototransistors, 4-terminal measurements lead to considerably higher values for the *EQE*, i.e. larger by a factor of 344% when compared to the more conventional 2-terminal measurements widely adopted by the research community.

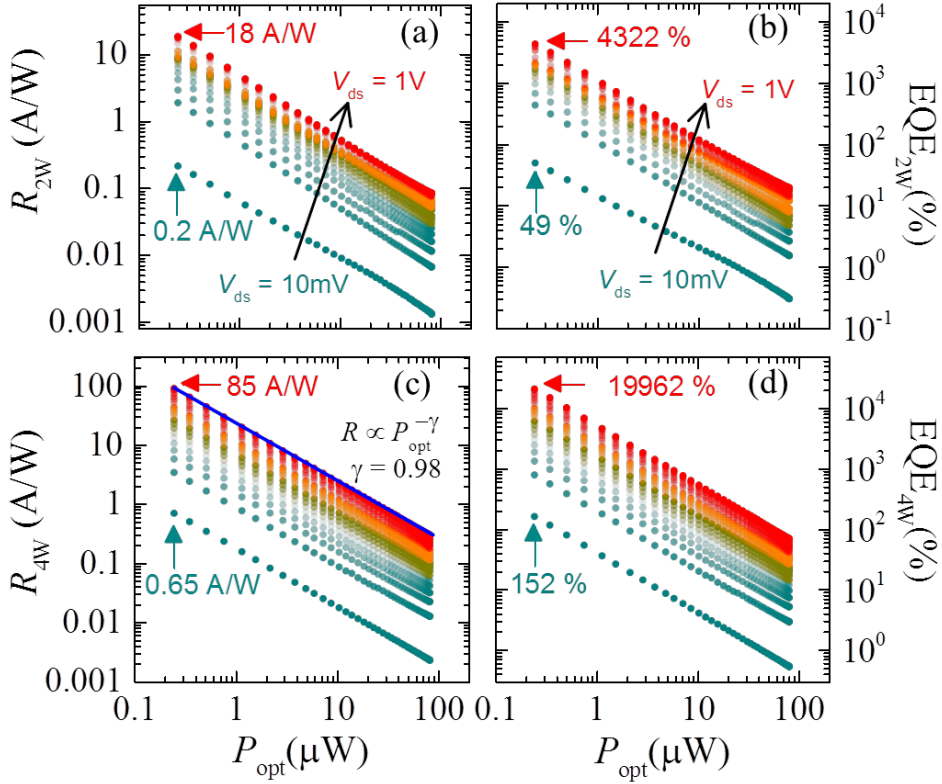

**Fig. 4** (a) and (c) Logarithmic plot of the responsivity for a few layered $WSe_2$ FET as a function of the illumination power for a 2- and a 4-terminal configuration of measurements, respectively. (b) and (d) *EQE* values extracted from equation (2). Both *R* and *EQE* are plotted for several values of the applied $V_{ds}$, i.e. ranging from 10 mV to 1 V. The blue solid line in (c) shows a linear fit of *R* as a function of $P_{opt}$ taken under $V_{ds}$ = 1 V to extract the exponent $\gamma$. $R_{2W}$ and $EQE_{2W}$ correspond to the responsivities and external quantum efficiencies measured with a 2-terminal configuration while $R_{4W}$ and $EQE_{4W}$ correspond to the ones collected with a 4-terminal configuration.

The decrease in the *EQE* value with increasing excitation power is likely due to electron-hole recombination processes in $WSe_2$ crystals [25]. The extremely high *EQE* values obtained from our few-layered $WSe_2$ phototransistor are due to the high *R* and large surface to volume ratio of 2D layered $WSe_2$ crystals and can be explained through a few scenarios [23]: *EQE* values exceeding 100% would seem to suggest the generation of multiple electron-hole pairs from a single absorbed photon, though our excitation frequency is too low for this to be expected to occur. High responsivity might also be achievable by preventing electron-hole recombination processes [23, 26-28]. According to Li *et al.*, [27] and Ulanganathan *et al.*, [23] charge traps may localize one type of photogenerated charge, thus increasing the lifetime of the free carrier which might circulate multiple times between the channel and sourcemeter before recombination thus contributing to a higher responsivity [28] by creating an effective gain mechanism. Since we illuminate the entire area of sample, including the contacts, it is also quite likely that the photons provide enough kinetic energy to charge carriers to promote them across the Schottky barrier thus considerably increasing the photocurrent. If this is confirmed to be the case, it would indicate that one could observe a dramatic increase in performance from FETs based on TMDs by just achieving ohmic contacts. In this report our aim was to study the intrinsic photoconductivity using a multi-terminal configuration of contacts under a single wavelength of laser excitation, or $\lambda = 532$ nm. Here we do not report wavelength dependent photoconductivity. We chose $\lambda = 532$ nm laser source because a previous report [12] suggested maximum photoconductivity at this wavelength for monolayer $WSe_2$. In effect, Zhang *et al.* [12] reported wavelength dependent photoconductivity and absorption on monolayer $WSe_2$ finding that they reach their maximum at $\lambda \sim 532$ nm incident wavelength. The responsivity was observed to decrease as they increased the wavelength. Thus, we conducted a detailed intrinsic photoconductivity study using a multi-terminal configuration under a 532 nm laser excitation.

Under illumination the measured total current is due to several contributions such as thermally excited minority charge carrier (observed under dark conditions or without laser illumination), photovoltaic/photogenerated current due to electron-hole pair generation under illumination, and the photothermoelectric (PTE) effect under illumination on the metal-semiconductor interface [29]. By illuminating the Ti/Au metal and $WSe_2$ interface under laser source, the temperature of the metallic contacts becomes slightly higher than that of the $WSe_2$ semiconductor due to their different Seebeck coefficients. This temperature difference results in PTE current across the metal-semiconductor junctions, which allows the carrier to flow from the semiconductor (cold) to metal (hot) while current also flows from the metallic contact to semiconductor channel at their interface [29]. In our experiment, the area of illuminated light (~32 μm diameter spot size ≈ 804 μm² area) is much larger than the device geometry (≤ 100 μm²), thus the PTE current generated from our device from both contacts should balance each other as the two PTE generated currents flow in opposite directions and therefore should not contribute to the total current measured.

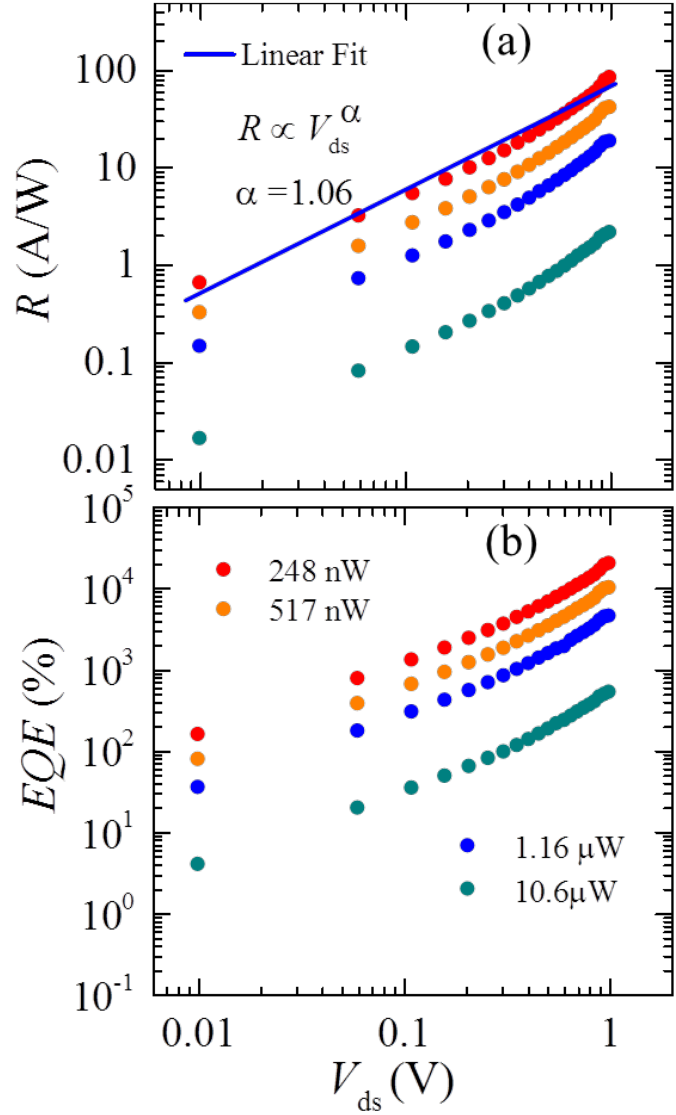

**Fig. 5** (a) Logarithmic plot of the responsivity (*R*) as a function of $V_{ds}$ for several values of the laser power and for a fixed $V_{bg}$ = 10 V when measured using a 4-terminal configuration. (Blue solid line) Linear-fit of *R* as a function of $V_{ds}$ at $P_{opt}$ = 248 nW. (b) *EQE* as a function of $V_{ds}$ for several values of the laser power. Both *R* and *EQE* increase considerably as functions of $V_{ds}$

To study how $R$ and $EQE$ depend on the bias voltage $V_{ds}$, we extracted the responsivities and external quantum efficiencies as functions of $V_{ds}$ from the 4-terminal measurements in Figs. 4 (c) and 4 (d) and plotted them in Fig. 5 as functions of $V_{ds}$ for several values of the laser power. Figure 5 (a) shows $R$ as a function of $V_{ds}$ under laser powers $P_{opt}$ ranging from 248 nW to 10.6 µW. In our study, $R$ increases linearly as a function of the applied $V_{ds}$ which varies from 10 mV up to 1 V. We also fitted $R$ as a function of $V_{ds}$ under $P_{opt}$ = 248 nW to a power law ($R \propto V^{\alpha}$), and the extracted value of $\alpha$ was ~1.06, which confirmed the observed high photoconducting gain in our few layered WSe$_2$ device. Figure 5 (b) displays the $EQE$ as a function of $V_{ds}$, as extracted from Fig. 4 (d), for several fixed values of the laser excitation power. For $V_{ds}$ ranging from 10 mV to 1 V, $EQE$ scales linearly with $V_{ds}$. The highest observed $EQE$ value is $2 \times 10^4$% for $V_{ds}$ = 1 V, which is considerably higher than the value observed using 2-terminals.

An extremely high responsivity of ~880 A/W was reported for monolayer MoS$_2$ under $V_{bg}$ = -70 V, $V_{ds}$ = 8 V, and with $P_{opt}$ = 150 pW[5]. This value decreased to < 4 A/W under $P_{opt}$ = 248 nW. Notice that this $R$ value should be much smaller than 4 A/W when $V_{ds}$ is extrapolated to 1V, and hence it would be orders of magnitude smaller than the value measured by us in few-layered WSe$_2$ under the same conditions. A similarly high responsivity $R$ ~ 95 A/W was reported for few layered MoSe$_2$ phototransistors under $P_{opt}$ = 15 nW [30] and $V_{ds}$ = 8 V, which is a much higher $V_{ds}$ value than the one used to measure our WSe$_2$ phototransistors, where we achieved similar responsivities but for $V_{ds}$ = 1 V and $P_{opt}$ = 248 nW. For monolayer WSe$_2$ under $V_{bg}$ = 8 V and $V_{ds}$ = 2 V, the previously reported responsivity was just ~0.02 A/W [12]. By increasing the number of layers, we obtain responsivities of 85 A/W under $V_{bg}$ = 10 V which is three orders of magnitude higher than those values observed in monolayers of WSe$_2$.

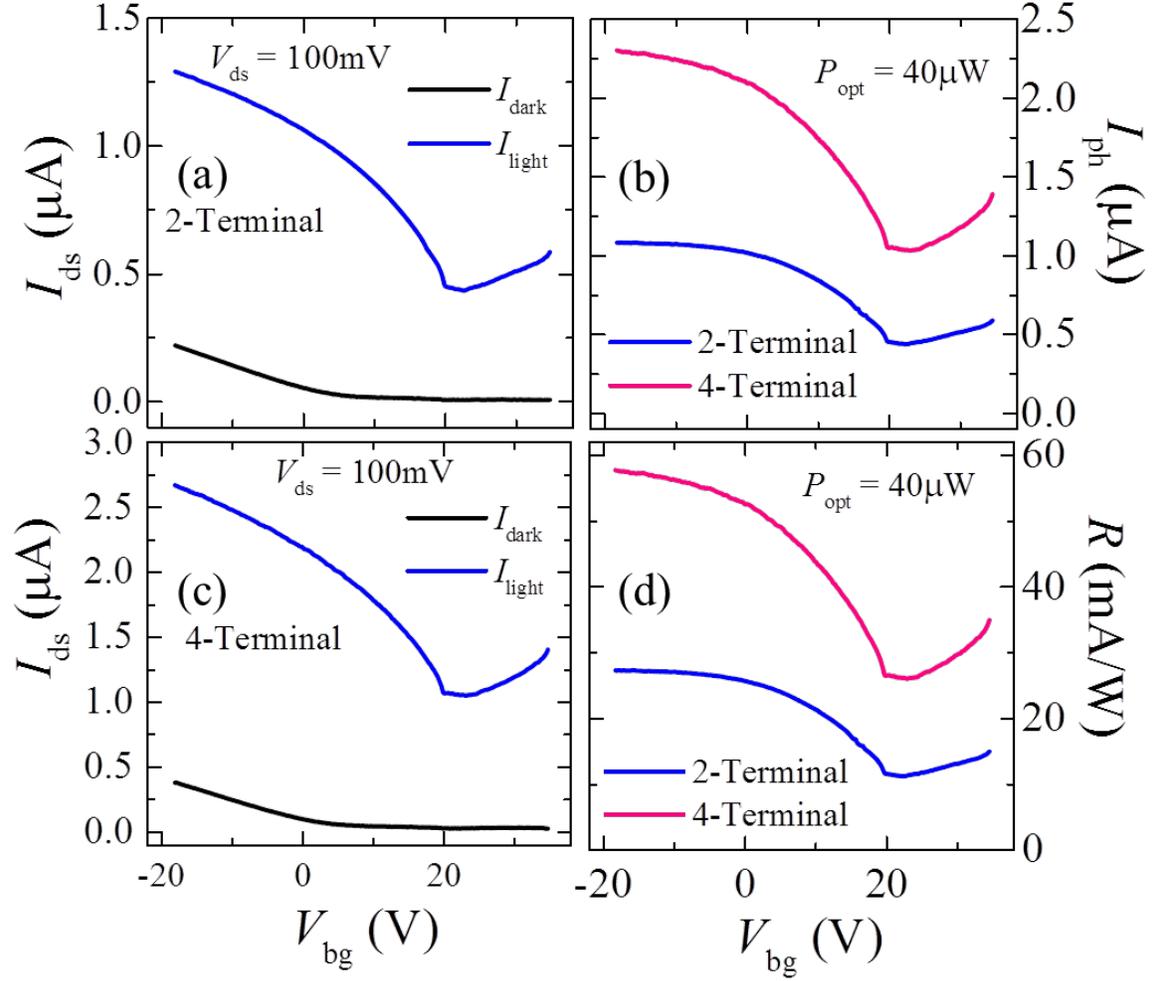

**Fig. 6**. (a) and (c) Drain to source current as a function of the back gate voltage $V_{bg}$ for a bias voltage $V_{ds}$ = 100 mV in absence of laser illumination (black line) and under laser illumination at $P_{opt}$ = 40 μW (blue line) measured with a 2- and a 4-terminal configuration, respectively. (b) Photocurrent ($I_{ph} = I_{ds} – I_{dark}$) as a function of the back gate voltage measure with 2- (blue line) and 4-terminal (magenta line) configuration of contacts, extracted from (a). (d) Responsivity as a function of $V_{bg}$ for both 2- and 4-terminal configurations.

We fabricated several devices and studied their phototransport behavior using both 2- and 4-terminal configurations. All of them showed similar trends, namely much higher responsivites and *EQE* values when using a 4-terminal configuration relative to the values obtained when using the 2-terminal one. To evaluate the gate-dependent optical response from WSe$_2$ phototransistors, we measured the photoconductivity at a fixed $V_{ds}$ and fixed laser excitation ($P_{opt}$ = 40 μW) using both configurations of contacts. Figure 6 shows such a comparison but measured from a much thinner WSe$_2$ crystal relative to the ~10 layers discussed above. The optical image of this ~3 to 4 layers WSe$_2$ sample is shown in the supplementary Fig. S4 (a). Its field-effect transport properties are displayed in the supplementary Fig. S4 (b). The two-terminal field-effect *hole*-mobility of this thinner device was 11 cm$^2$/Vs (5 cm$^2$/Vs in four-terminal configuration). Here, we study the phototransport behavior as a function of the applied gate-voltage under a fixed $V_{ds}$ = 100 mV and a laser power $P_{opt}$ = 40 µW.

Figure 6 displays a detailed study of the gate-dependent optical response from the thin-WSe$_2$ phototransistor fabricated on a Si/SiO$_2$ substrate with Ti/Au contacts as used in the previous devices. The black lines in Figs. 6 (a) and 6 (c) depict the drain-source current without illumination ($I_{dark}$) as a function of the applied gate voltage ($V_{bg}$) and for $V_{ds}$ = 100 mV in 2- and 4-terminal configurations, respectively. In contrast, the blue line represents $I_{ds}$ as a function of the gate voltage under a fixed laser illumination power $P_{opt}$ = 40 µW. This gate dependent response shows a nonlinear photocurrent response over the entire range of gate voltages presented here. We observed slightly higher photocurrents in the ON state of the transistor than in its OFF state. The photogenerated current $I_{ph}$ (= $I_{light}$ − $I_{dark}$) is displayed in Fig. 6 (b) for both 2- (blue line) and 4-terminal (magenta line) configurations, clearly showing a much higher photocurrent for the 4-terminal one and as seen in the previous device discussed above. The photogenerated current measured in a 4-terminal configuration under $V_{bg}$ = -18 V and under $V_{bg}$ = 10 V are respectively 114% and 108% higher than the 2-terminal ones. This difference is smaller than the one observed on the 10-layered device described above. This difference could occur for several possible reasons. It could occur due to the higher contact resistance to a thinner flake. Additionally, a smaller number of electron-hole pairs could be generated due to a weaker interaction with light in the thinner material. Finally, the thinner sample is also more susceptible to phonon scattering or a larger substrate interaction.

The responsivity of the thinner device as a function of the gate voltage is displayed in Fig. 6 (d) for both 2- (blue) and 4-terminal (magenta) configurations. The responsivity of the 10-layered sample under $V_{bg}$ = 10 V and $V_{ds}$ = 100 mV is 181 mA/W when measured with a 4-terminal configuration, [see, Fig. 4 (c)] while the corresponding value for the thin-layered WSe$_2$ crystal is 43.5 mA/W. This clearly shows that the responsivity increases considerably with the thickness of the device, which agrees well with previous reports on MoS$_2$ phototransistors [31]. We additionally provide the $I_{ds}$ vs $V_{ds}$ and photocurrent dependence on applied optical power of a third device in supplemental information Figures S5 and S6. This device had WSe$_2$ with ~7-8 layers, and was intermediate in performance to the two devices we already reported. In the third device, we measured a 4-terminal responsivity of ~4.8 A/W with $V_{ds}$ = 0.5 V and $V_{bg}$ = 0 V at $P_{opt}$ = 0.035 µW, which was a 170% increase over the corresponding 2-terminal responsivity. At $P_{opt}$ = 0.35 µW, the increase in responsivity using a 4-terminal configuration was 250%. Using mechanically exfoliated WSe$_2$ naturally leads to a variability in absolute device performance, but a significant increase in the 4-terminal responsivity over the 2-terminal was consistently observed. The thicker (~10 layers) and thinner (~3-4 layers) devices represent an upper and lower bound, respectively, for the performance of the devices we tested.

**Conclusion**

We reported a comparison between the phototransport properties of few-layered WSe$_2$ phototransistors when measured in a conventional two-terminal and in a multi-terminal configuration. We studied in great detail, the photoconductivity, responsivity, and the external quantum efficiency as a function of the laser excitation power and bias voltage using both configurations. Our investigations clearly indicate that by minimizing the role of the resistance of the contacts one can observe much higher responsivities in transition metal dichalcogenides. For example, our photoconductivity study on transistors composed of just ~10 atomic layers of WSe$_2$ display a 370% higher photocurrent and concomitant external quantum efficiency when compared to the values obtained from a conventional two-terminal measurement. This indicates that the intrinsic photoconductivity of transition metal dichalcogenides is considerably higher than the values reported so far. The correct evaluation of their optoelectronic properties should reveal their true potential for applications, which our study suggests would seem to be considerably higher than previously considered. We hope that our study will encourage the community to expose the nearly intrinsic properties of these compounds, with a perspective on optoelectronic applications, until a solution is found to produce nearly ohmic contacts.

**Materials and Methods**
High quality WSe$_2$ single crystals were synthesized by the chemical vapor transport technique using Se or Iodine as the transport agents. Previously we checked the quality of the synthesized crystals by EDX and Raman spectroscopy [20]. Bulk WSe$_2$ single crystals were exfoliated by using the micro-exfoliation technique. Subsequently the exfoliated thin layers of WSe$_2$ were transferred onto a clean SiO$_2$ *p*-doped Si substrate and identified under an optical microscope. The thickness of the flakes was determined *via* atomic force microscope techniques.

**Fabrication of field-effect transistors**
We used standard electron-beam lithography techniques to pattern the contacts, and then an electron beam evaporator at $10^{-7}$ torr to deposit the Ti/Au (~5/85 nm) contacts. The FETs were annealed at 300$^{\circ}$C for 3 hours in forming gas followed by vacuum ($10^{-7}$ torr) annealing at 120$^{\circ}$C for 24 hours.

**Photoconductivity measurements**
Photoconductivity measurements were performed in a home built optical microscope which permits white light illumination with a lamp or with a 532 nm laser source. A small CCD camera was also used to image the samples and to facilitate the alignment of the laser spot on the active area of the sample. Photoconductivity measurements were taken under a constant back gate and source-to-drain voltages while the incident laser power varied from 0 to 500 μW by using a variable attenuator. A mechanical shutter was used to block the laser light in order to collect the dark current of the FET. Illumination was concentrated onto the sample by using a 10x objective, which produced a laser spot size of 35 μm in diameter. The diameter was measured by focusing the laser spot onto a pre-patterned grid on a Si wafer. The incident laser power was calibrated with a Si-based optical power meter prior to the experiments. White light illumination was not used during data collection.

## Author contribution statement

NRP, SAM, and LB conceived the project. DR synthesized the $WSe_2$ single-crystals through the CVT technique. NRP and CP fabricated the devices. CG, JH, and SAM built the experimental setup for optical measurement. CG, JH, NRP, SAM measured the optical properties, NRP analysed the data and NRP, SAM and LB wrote the manuscript with the input of all co-authors.

## Acknowledgements

NRP, DR and LB acknowledge supported from U. S. Army Research Office through the MURI grant W911NF-11-10362. SAM acknowledges support from NSF through DMR-1229217. The NHMFL is supported by NSF-DMR-0084173 and the state of Florida.

## Notes and references

Corresponding authors email: *mcgill@magnet.fsu.edu, pradhan@magnet.fsu.edu

Supplementary Information (SI) available: [Supplementary information containing $I_{ds}$ vs $P_{opt}$, $I_{ds}$ vs $V_{ds}$, and optical image of thinner device is available].

# Electronic Supporting information for manuscript "Photoconductivity of few-layered p-WSe$_2$ phototransistors *via* multi-terminal measurements"

Nihar R. Pradhan [a,*], Carlos Garcia [a,b], Joshua Holleman [a,b], Daniel Rhodes [a,b], Chason Parker [a,c], Saikat Talapatra [d], Mauricio Terrones [e], Luis Balicas [a] and Stephen A. McGill [a,*]

[a]National High Magnetic Field Laboratory, Florida State University, Tallahassee, FL 32310, USA
[b]Department of Physics, Florida State University, Tallahassee, FL 32306, USA
[c]Leon High School, Tallahassee, FL 32308, USA
[d]Department of Physics, Southern Illinois University, Carbondale, IL 62901, USA
[e]Department of Physics, Pennsylvania State University, PA 16802, USA

1. **Photocurrent as a function of applied laser power excitations with several applied drain-source voltage from 60 mV to 1 V for few-layered WSe$_2$ field effect transistors.**

Figure S1 displays the photocurrent as a function of applied laser power from 532 nm laser source. Photogenerated current was extracted by subtracting the dark current (without light illumination) from the current measured with light incident on the sample. Photocurrent displayed in Fig. S1 is measured at applied $V_{ds}$ = 60 mV to 1 V from bottom (dark cyan) to top (red) of the data plots. The results show the photogenerated current varies from 0.5 μA at applied drain-source bias $V_{ds}$ = 60 mV to 7 μA at $V_{ds}$ = 1 V. The photogenerated current looks less power dependent at lower applied $V_{ds}$ but became more dependent at higher $V_{ds}$.

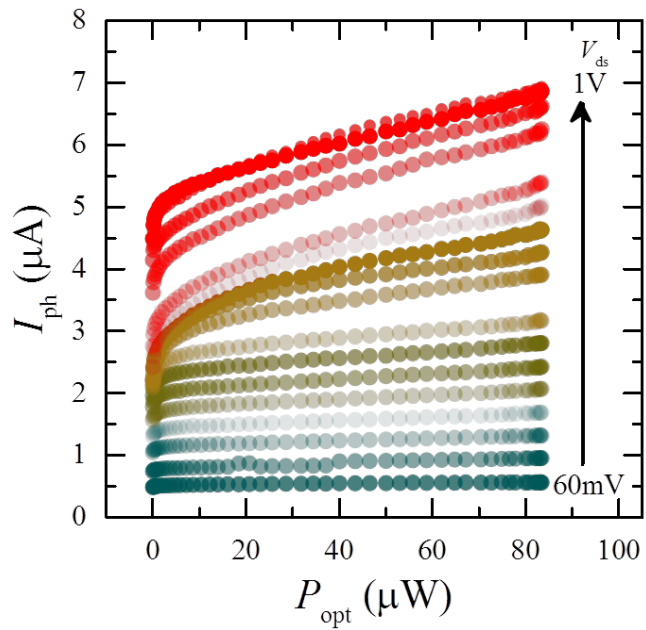

**Figure S1.** Photogenerated current $I_{ph}$ (=$I_{light} - I_{dark}$) as a function of laser excitation power incident on the sample. The photocurrent was measured at several drain-to-source voltage from 60 mV to 1 V at back gate voltage $V_{bg}$ = 10 V.

2. $I_{ds}$ vs $V_{ds}$ and $I_{ds}$ vs $V_{bg}$ of device #1 presented in semilogarithmic scale.

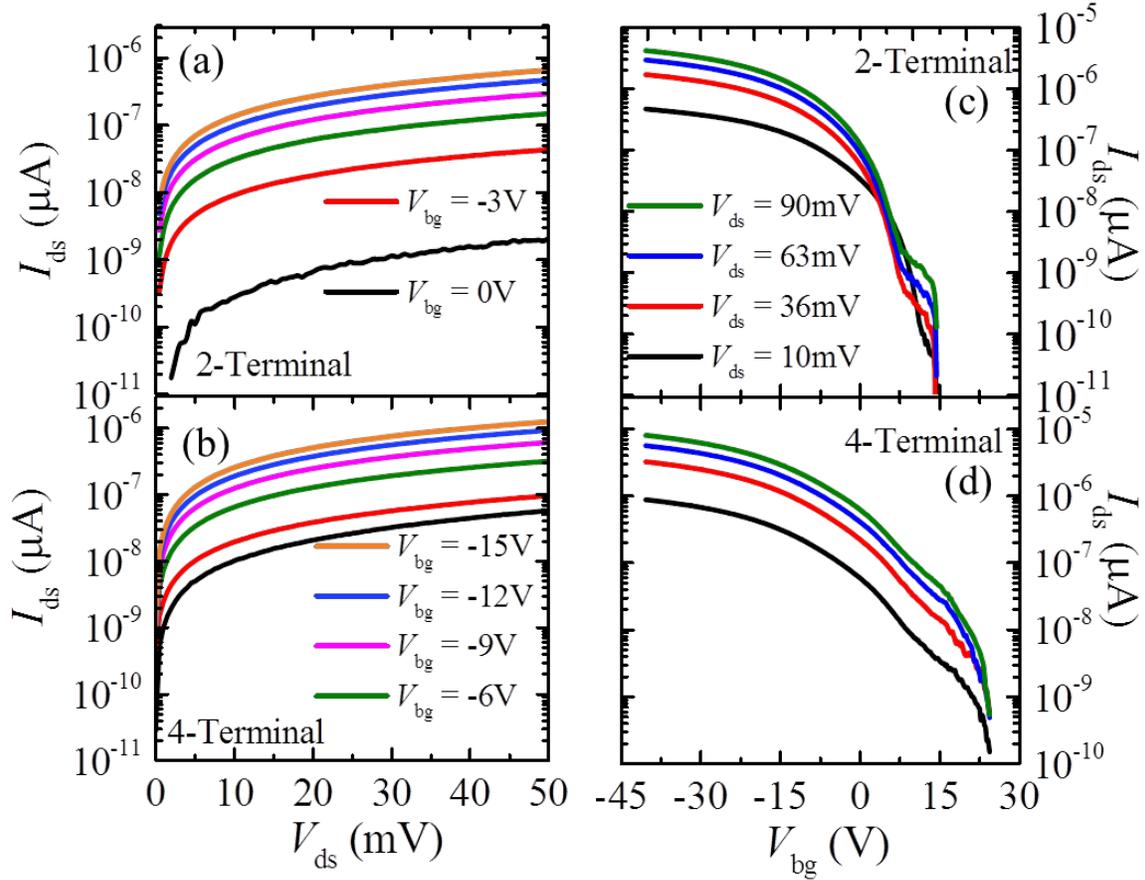

**Figure S2.** Shows the drain-source current as a function of drain to source and drain to back gate voltage in semi-logartmic scale measured in 2-terminal and 4-terminal configurations. The linear scaled data is presented in Fig. 2 of main text.

### 3. Field-effect mobility and contact resistance of few-layered WSe$_2$.

Figure S2 (a) displays the $I_{ds}$ vs $V_{bg}$ at $V_{ds}$ = 90mV. We extracted the field-effect mobility from the slope taken from the linear part of $I_{ds}$ vs $V_{ds}$ graph by using MOSFET 2-terminal transconductance formula $\mu_{FE} = \frac{L}{w}\left(\frac{dI_{ds}}{dV_{bg}}\right)\frac{1}{C_i}\frac{1}{V_{ds}}$, where $L$ = length of the channel, $w$ = width of the channel, $C_i$ = capacitance per unit area. The extracted field-effect mobility is ~220 cm$^2$/Vs. Figure S2 (b) show the contact resistance of Ti contact on few-layered WSe$_2$, extracted from 2-

Terminal and 4-Terminal measurements, $I_{ds}$ vs $V_{ds}$ graph shown in Figure 2 (a) & (b) in the main text. The ON/OFF current ratio of the transistor is exceeding $10^6$ shown in Fig. S2 (b).

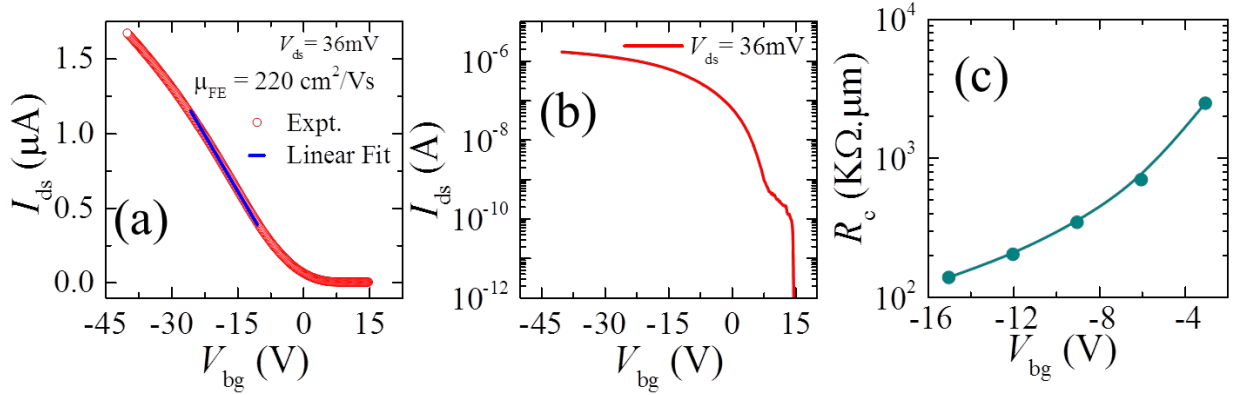

**Figure S3.** (a) Drain-to-source current ($I_{ds}$) as a function of back gate voltage ($V_{bg}$) of few-layered WSe$_2$ field-effect transistor at fixed $V_{ds}$ = 36mV. (b) displays the logarithmic plot of $I_{ds}$ vs $V_{ds}$ as a function of $V_{bg}$ indicates the FET shows high ON/OFF current ratio >$10^6$. (c) Contact resistance ($R_c$) as a function of back gate voltage.

## 4. Optical micrograph image and transport graph of a second thin layered WSe$_2$ sample.

Figure S3 displays the second four layers WSe$_2$ FET with multi-contacts for 2-terminal and 4-terminal measurement. The field-effect tranport graph is shown in Fig. S3 (b), which displays the drain-source currnet as a function of back gate voltage at applied drain-source voltage $V_{ds}$ = 100mV. The calculated field-effect mobility obtained from this device is 11 cm$^2$/Vs from 2-terminal measurement. This mobility is much less than the 10-layered WSe$_2$ sample presented in the main text. This may be due to higher contact resistance between Ti-metal contact and thin layers of WSe$_2$ crystal. It could also be due to higher phonon scattering and large substrate interaction of atoms in thin crystal than the thicker layers.

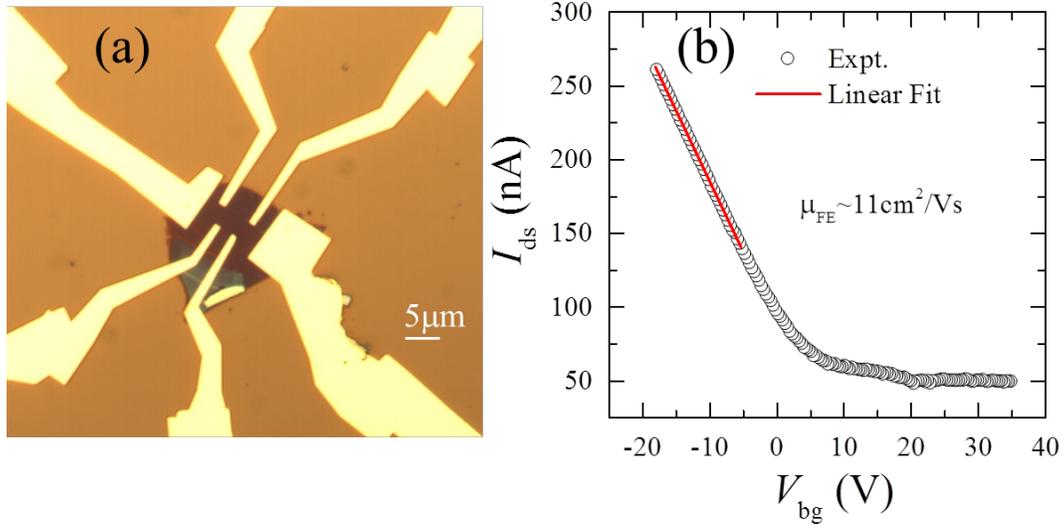

**Figure S4.** (a) Optical micrograph image of second thin-layered WSe$_2$ field-effect transistor with multi-terminal configuration. The length and width of the device is 10 μm and 8 μm respectively. (b) The drain-to-source current as a function of backgate voltage at constant $V_{ds}$ = 100 mV is displayed. The red line is the linear fit to the experimental data.

**5. $I_{ds}$ vs $V_{ds}$ characteristic and optical image of device #3**

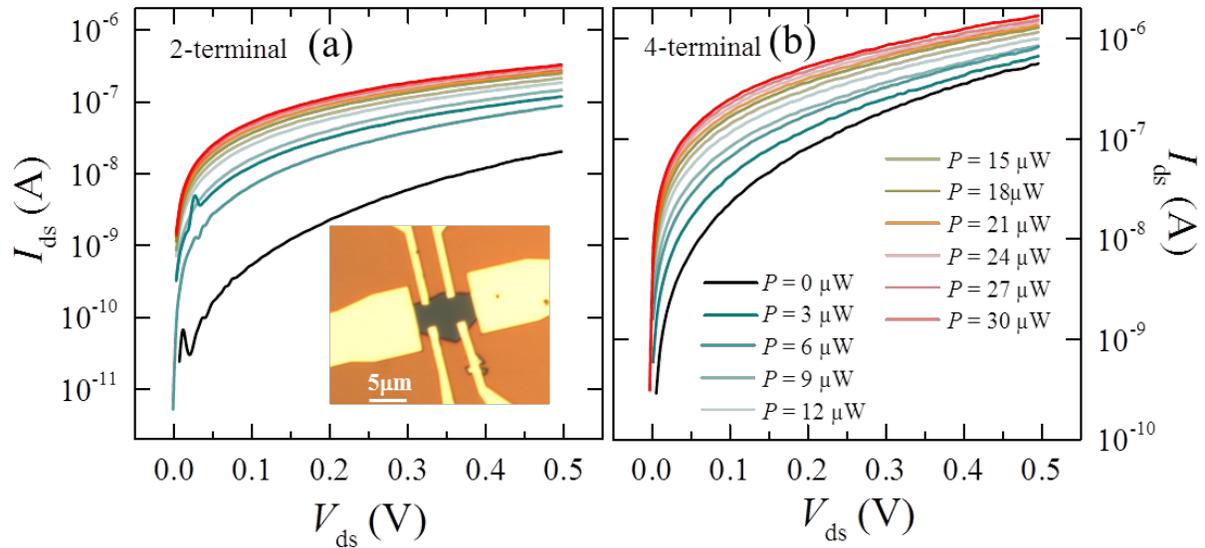

**Figure S5:** $I_{ds}$ vs $V_{ds}$ with several applied laser powers in 2-terminal (a) and 4-terminal (b) configurations at $V_{bg}$ = 0V. Inset in (a) shows the optical image of the WSe$_2$ FET of ~ 5.8 nm thick (7-8 layers).

## 6. Photocurrent and responsivity of device #3

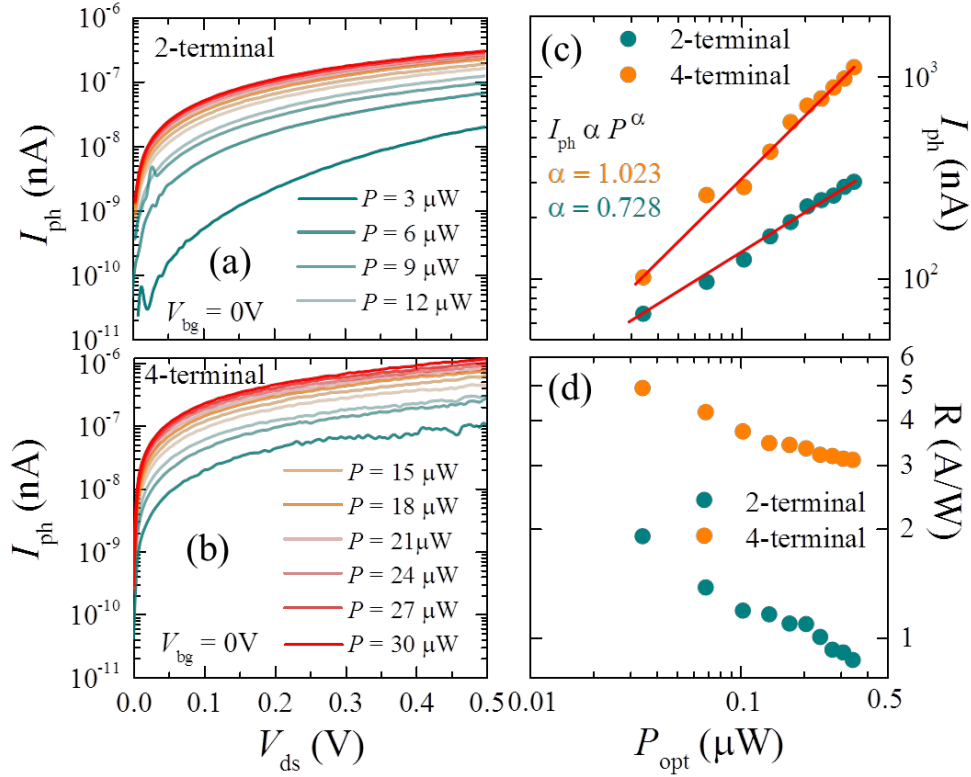

**Figure S6.** (a) and (b) show photocurrent ($I_{ph} = I_{ds} - I_{dark}$) as a function of drain-source voltage ($V_{ds}$), measured in 2-terminal and 4-terminal configurations, respectively, at several applied laser powers and at $V_{bg} = 0$V. (c) displays the photocurrent as a function of applied optical power in logarithmic scale. The red lines are the power law fit $I_{ph} \propto P^{\alpha}$ for both 2-terminal and 4-terminal configurations. 4-terminal measurement shows higher photocurrent and exponent ($\alpha$) compared to 2-terminal configurations. The photoresponsivity ($R$) as a function of applied laser power is shown in Figure (d). The value of $R$ at $P_{opt} = 0.4$ μW measured at 4-terminal, is 250% increase over 2-terminal measurement.